\documentstyle[prbbib,aps,prb,epsfig,twocolumn]{revtex}

\begin{document}
% \draft command makes pacs numbers print
\draft

\wideabs{

\title{Stimulated emission and ultrafast carrier relaxation in InGaN multiple quantum wells}

\author{\"{U}mit \"{O}zg\"{u}r, and Henry O. Everitt$^{\rm a)}$
\footnotetext{$^{\rm a)}$Electronic mail: everitt@aro.arl.army.mil} }

\address{Department of Physics, Duke University, Durham, NC 27708}

\author{Stacia Keller and Steven P. DenBaars}
\address{Department of Electrical Engineering and Materials Science, University of California, Santa Barbara, CA}

\date{\today}

\maketitle

\begin{abstract}
Stimulated emission (SE) was measured from two InGaN multiple quantum well
(MQW) laser structures with different In compositions. SE threshold power
densities (I$_{th}$) increased with increasing QW depth ($x$).
Time-resolved differential transmission measurements mapped the carrier
relaxation mechanisms and explained the dependence of I$_{th}$ on $x$.
Carriers are captured from the barriers to the QWs in $<$ 1 ps, while
carrier recombination rates increased with increasing $x$. For excitation
above I$_{th}$ an additional, fast relaxation mechanism appears due to the
loss of carriers in the barriers through a cascaded refilling of the QW
state undergoing SE. The increased material inhomogeneity with increasing
$x$ provides additional relaxation channels outside the cascaded refilling
process, removing carriers from the SE process and increasing I$_{th}$.

\end{abstract}


}

Recent advances in the growth of group-III nitride heterostructures have made
it possible to manufacture efficient green, blue, and ultraviolet emitters
and detectors.\cite{NakamuraBLD1997,NakamuraJJAP1997-2,NakamuraJJAP95}
Carrier dynamics in such device structures are beginning to be understood
through the use of high power, short pulse, regenerative and optical
parametric
amplifiers.\cite{OzgurAPL2000,OzgurPRB,KawakamiAPL2000,SatakePRB1999} In
this paper, stimulated emission (SE) in InGaN multiple quantum well (MQW)
laser structures with differing QW depths is explored. SE threshold power
densities increase with increasing QW depth, and ultrafast measurements of
carrier dynamics reveal why.

Two different InGaN MQW samples, differing primarily by QW In composition
($x$), were grown on c-plane sapphire in a modified two flow metalorganic
chemical vapor deposition (MOCVD) reactor.\cite{KellerJCG1998} Both
samples have $\sim$2 $\mu$m thick GaN:Si base layers. The 12\% sample
consists of 10 periods of 8.2 nm/3.4 nm
In$_{0.07}$Ga$_{0.93}$N:Si/In$_{0.12}$Ga$_{0.88}$N QWs and is capped with
a 100 nm thick GaN:Si layer. The 23\% sample consists of 12 periods of 10
nm/3.5 nm In$_{0.03}$Ga$_{0.97}$N:Si/In$_{0.23}$Ga$_{0.77}$N QWs and is
capped with 16 nm thick Al$_{0.10}$Ga$_{0.90}$N and 40 nm thick GaN:Si
layers. The In compositions and the QW/barrier thicknesses are determined
by high resolution X-Ray measurements. Material inhomogeneities, which
arise from QW thickness fluctuations, compositional fluctuations, and In
phase separation, have previously been observed to increase with
increasing $x$ and will be shown to play an important role in carrier
relaxation and emission processes.\cite{ChichibuJVST1998,ChichibuMSEB1999}

\begin{figure}
\centerline{\resizebox{9cm}{!}{\hbox{\includegraphics {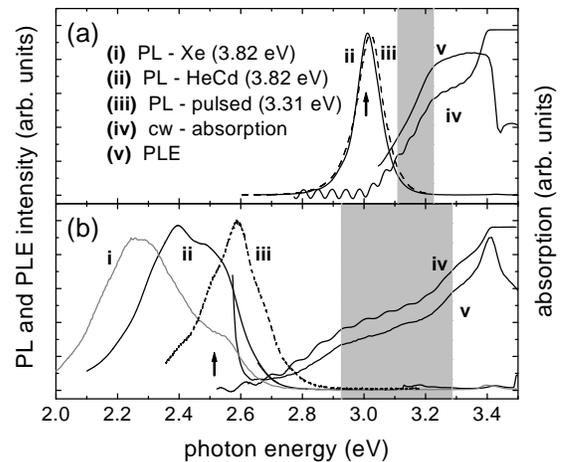}}}}
\caption{Room temperature cw-PL using a Xe lamp source at 325 nm (i) and a
HeCd laser (ii); time-integrated, below $I_{th}$, pulsed-PL (iii),
Cw-absorption (iv), and PLE (v) for the (a) 12\%, and (b) 23\% MQW
samples. The shaded regions show the states between the barrier and QW
band edges. Arrows show the detection energies for the PLE.} \label{PLPLE}
\end{figure}

Fig.~\ref{PLPLE} shows the cw-PL for 3.81 eV low intensity (300W Xe lamp)
and moderate intensity (25mW HeCd laser at 325 nm) excitations, and
time-integrated PL (TI-PL) with a high intensity ($\sim$15 $\mu$J/cm$^2$)
pulsed 3.31 eV excitation for both MQW samples. PL for the 12\% sample
shows a single peak at 3.01 eV for both Xe lamp (not shown) and HeCd
excitations. The 23\% sample shows at least two well-defined PL peaks for
Xe-Lamp and HeCd laser excitation, the higher energy of which (2.52 eV) is
assigned as the main PL from the QWs. The lower energy PL peak is
associated with impurity (yellow) emission. For both samples, TI-PL from
spontaneous emission (SPE) blueshifted for with increasing pump pulse
intensity (Fig.~\ref{PLPLE}) due to band filling and piezoelectric (PZE)
field screening by the increasing number of laser-induced
carriers.\cite{ChichibuJVST1998} The intensity-dependent blueshifts and PL
linewidths increased with increasing $x$, indicating a greater role of
inhomogeneities.

Cw-absorption measurements (Fig.~\ref{PLPLE}) revealed the 3D barrier
energies as 3.23 eV and 3.29 eV for the 12\% and 23\% samples,
respectively. Although the 12\% sample showed no remarkable QW band edge
absorption, a PLE edge at the confined QW energy is weakly visible at room
temperature (3.11 eV) and clearly visible at 77 K (3.14 eV). By contrast,
the QW band edge is clearly observed for the 23\% sample at 2.93 eV for
both cw-absorption and PLE measurements. More importantly, the PLE signal
was much broader for the 23\% sample and extended to localized states at
least 200 meV below the QW band edge. The broadening of the PLE, the
Stokes-like shift between the QW band edge and the emission energy, and
the number of localized states below the QW band edge clearly increased
with increasing $x$, another consequence of increased material
inhomogeneities and carrier localization.

For normal incidence pulsed excitation at moderate pump densities
($>$100$\mu$J/cm$^2$), stimulated emission (SE) is observed from the edge
parallel to the sample surface. The SE peaks for both samples appear near
the QW band edge, indicating that the SE originates from the lowest energy
QW confined states. Spectrally-integrated emission and linewidths were
measured as a function of the average pump density to obtain
I$_{th}$.\cite{SEbelowTh} For the 12\% sample, a broad SE peak appeared
slightly blueshifted (20 meV) from the PL peak. In contrast, the SE peak
for the 23\% sample (2.90 eV) appeared 200 meV  bluer than the PL peak.
The linewidth of the SE peak for the 23\% sample ($>$70 meV) is remarkably
larger than the linewidth for the 12\% sample ($\sim$30 meV), again a
consequence of material inhomogeneities.

SE thresholds (I$_{th}$) increased with $x$ (90, and 160 $\mu$J/cm$^2$ for
the 12\% and 23\% samples, respectively), contradicting earlier
observations that I$_{th}$ decreased with increasing QW
depth.\cite{KawakamiAPL2000,KwonAPL1999} In order to further investigate
this behavior, non-degenerate time-resolved differential transmission
(TRDT) spectroscopy was applied to measure carrier dynamics at room
temperature. A Ti:Sapphire laser-seeded Quantronix Titan regenerative
amplifier (RGA) with $\sim$1.8 mJ, $\sim$100 fs pulses at 800 nm was used.
Half of the RGA output power is used to pump a Quantronix TOPAS optical
parametric amplifier (OPA). The signal output from the tunable OPA is
frequency quadrupled and used as the pump in the TRDT experiment. The
other half of the RGA output is frequency doubled in a BBO crystal and
focused on a quartz cell filled with D$_2$O to generate a broadband
continuum probe. The pump beam is delayed with respect to the probe beam
using a retroreflector mounted on a 1 $\mu$m-resolution translational
stage. The probe, transmitted through the sample, is then collected by a
spectrometer with a liquid nitrogen cooled charge-coupled device camera.

\begin{figure}
\centerline{\resizebox{9cm}{!}{\hbox{\includegraphics {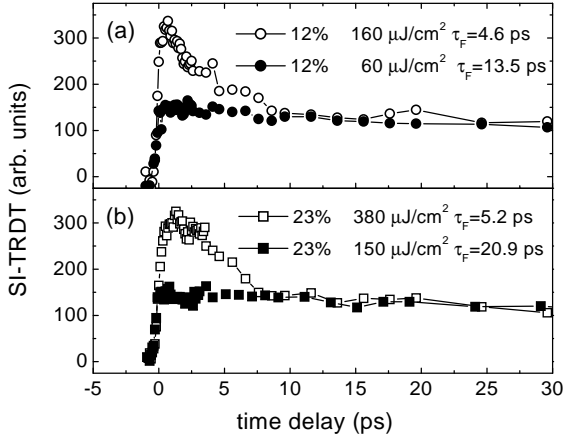}} }}
\caption{SI-TRDT for the (a) 12\% and (b) 23\% MQW samples, excited at
energies near the barrier band edge for densities above (open) and below
(filled) I$_{th}$.  Spectral integration occurred from the GaN band edge
(3.40 eV) to below the InGaN QWs (2.92 and 2.55 eV for the 12\% and 23\%
samples, respectively).} \label{IntegratedDT}
\end{figure}

To examine the relaxation of the total population of carriers, TRDT
signals are spectrally integrated (SI) over all energies where
photoexcited carriers were observed (Fig.~\ref{IntegratedDT}). For
excitation energies near the barrier band edge and power densities above
I$_{th}$, both samples show an initial fast decay followed by a much
slower relaxation. SI-TRDT data were fit by a bi-exponential decay
function, $Fe^{-t/\tau_F}+Se^{-t/\tau_S}$, where $\tau_F$ and $\tau_S$ are
the decay constants, and $F$ and $S$ are the magnitudes, for the fast and
slow decaying components, respectively. As the excitation density
decreased, $F$ decreased and $\tau_F$ slowed, while $S$ and $\tau_S$ were
unchanged (Fig.~\ref{IntegratedDT}). The fast decaying component in the
SI-TRDT data is caused by the accelerated relaxation of carriers through
SE.\cite{OzgurPRB,SatakePRB1999} After $\sim$10 ps the SE ends, leaving
only the much slower decaying component due to carrier recombination and
spontaneous emission.

\begin{figure}
\centerline{\resizebox{9cm}{!}{\hbox{\includegraphics
{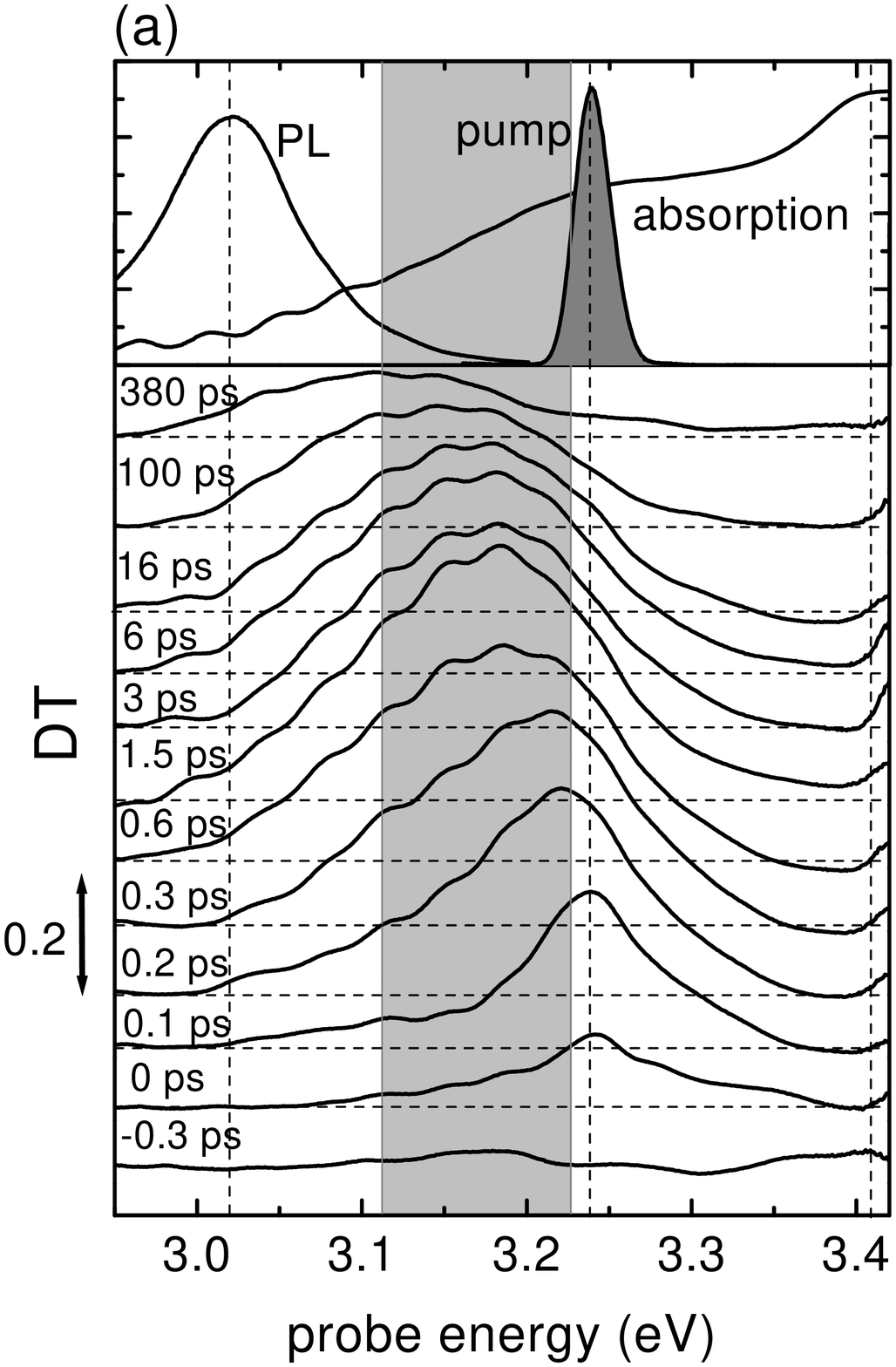}}\hbox{\includegraphics {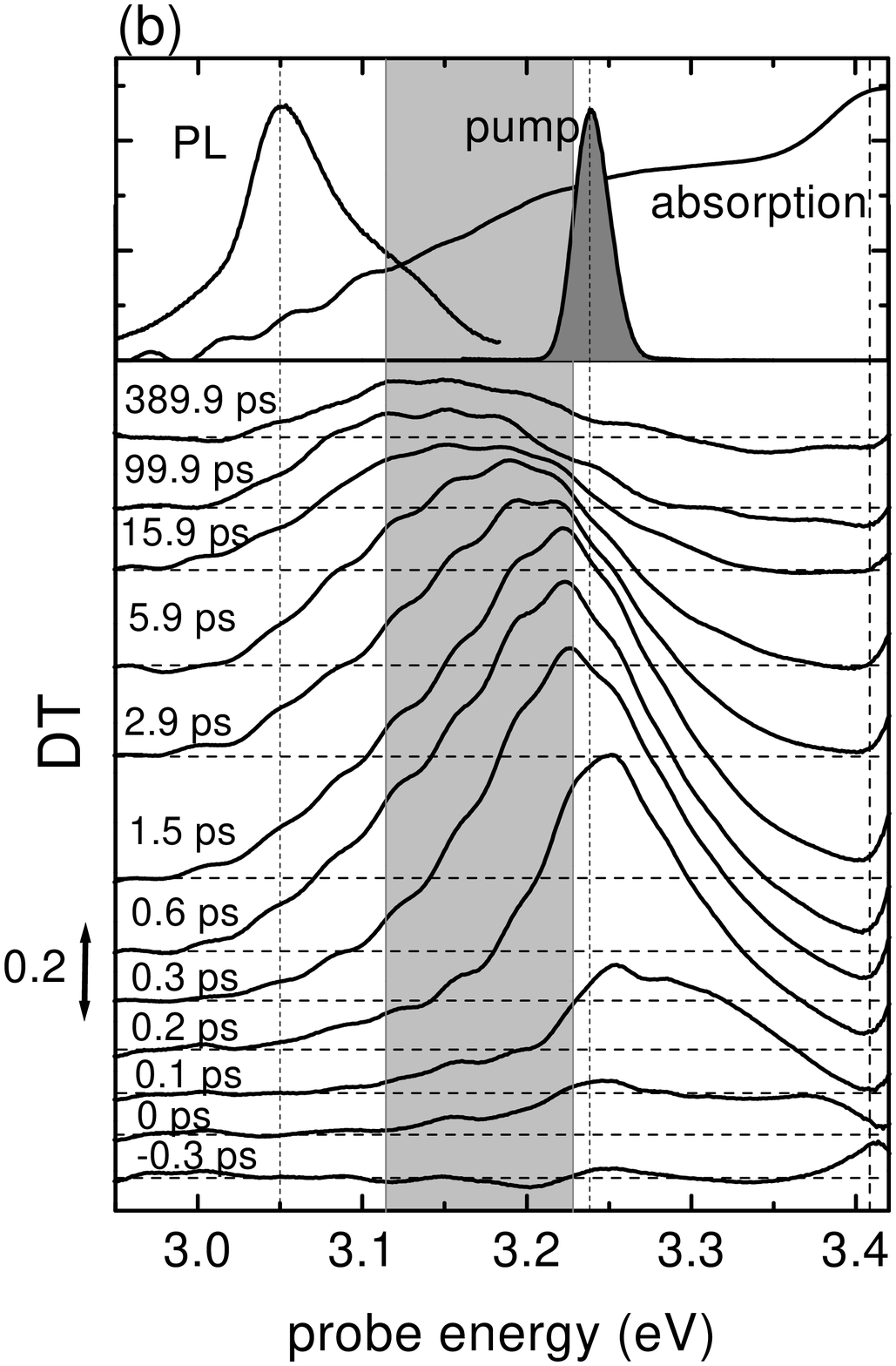}} }}
\centerline{\resizebox{9cm}{!}{\hbox{\includegraphics
{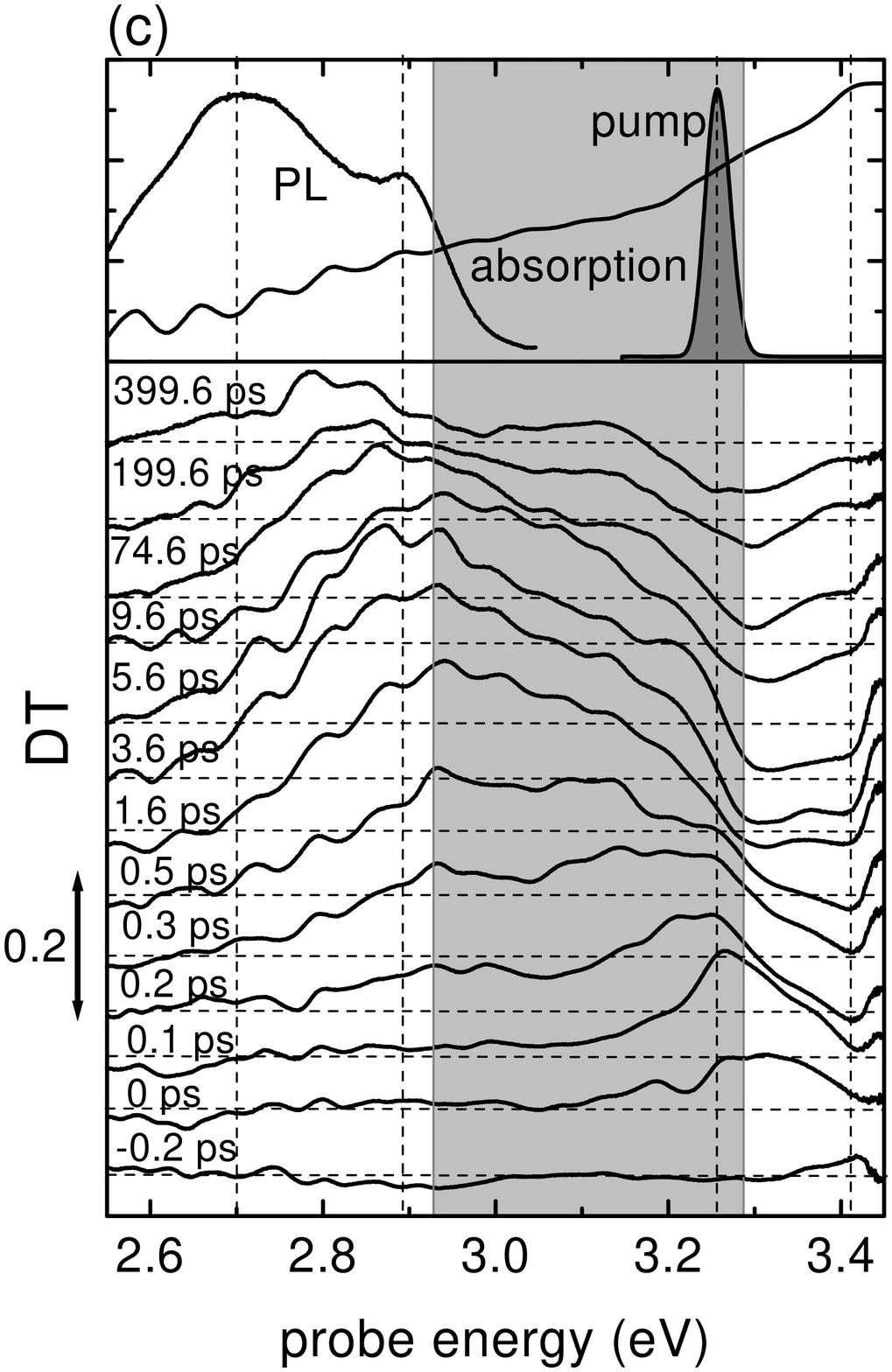}}\hbox{\includegraphics {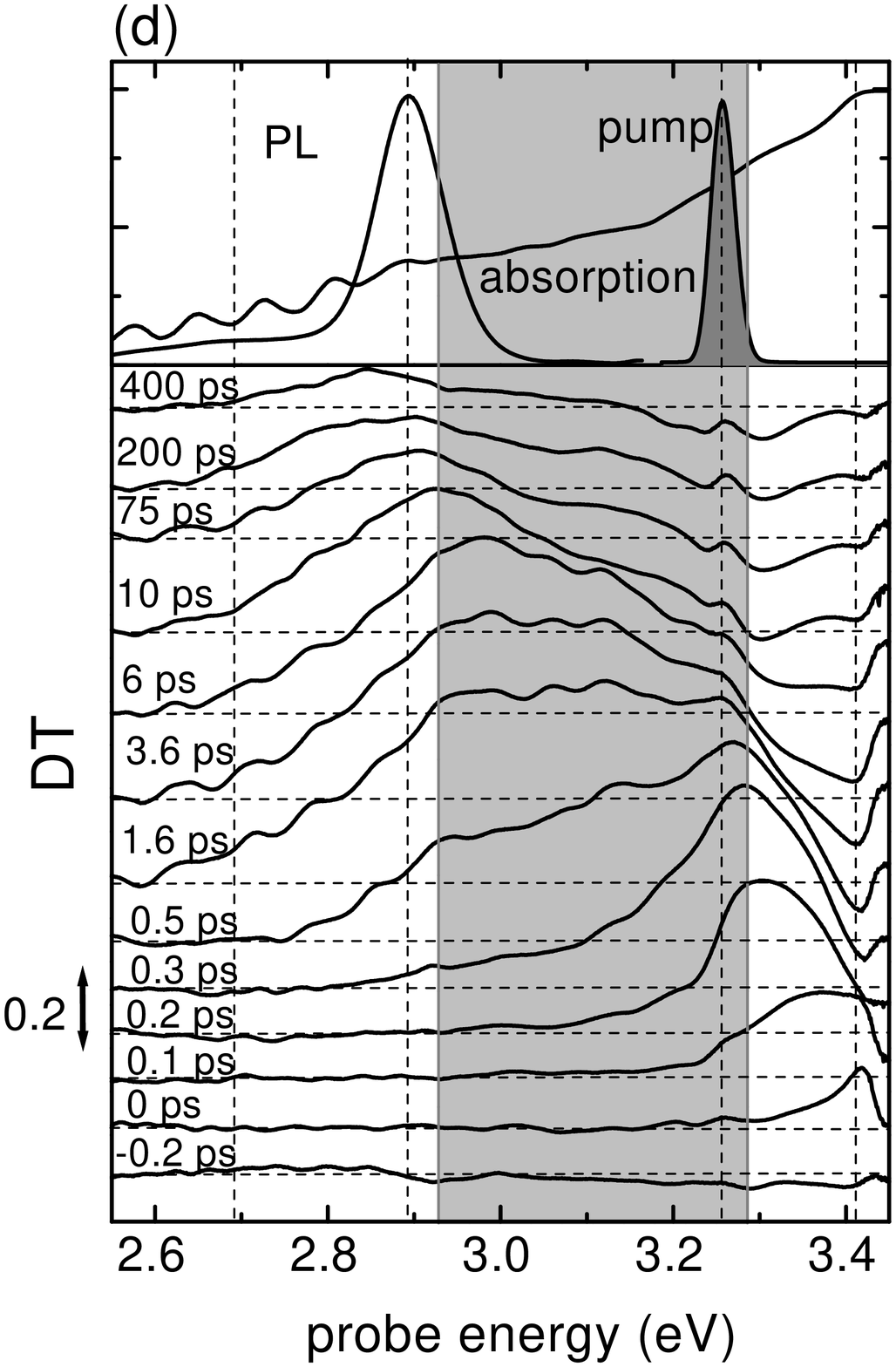}}}} \caption{SR-TRDT for (a)
below (60 $\mu$J/cm$^2$) and (b) above (160 $\mu$J/cm$^2$) I$_{th}$(12\%),
and (c) below (150 $\mu$J/cm$^2$) and (d) above (380 $\mu$J/cm$^2$)
I$_{th}$(23\%). Shaded regions show the states between the barrier and the
QW band edges.} \label{MQW-DT}
\end{figure}

Spectrally-resolved TRDT (SR-TRDT) data for the 12\% and 23\% samples
further elucidates the carrier redistribution and relaxation processes, as
shown in Fig.~\ref{MQW-DT}.\cite{StarkEffect} For excitation densities
below I$_{th}$ in the 12\% sample (I$_{th}$(12\%)), photoexcited carriers
relax to the barrier band edge in $<$1 ps, resulting in a broad
distribution (160 meV) of carriers in the QWs centered at energies that
increasingly redshift below the barriers with increasing time
(Fig.~\ref{MQW-DT}a). The arrival of carriers at the QW band edge (3.11
eV) $\sim$300 fs after carrier accumulation at the barriers (3.22 eV)
indicates the rate of carrier capture from the 3D barrier states into the
2D QW states. \cite{OzgurAPL2000,OzgurPRB} For excitation densities above
I$_{th}$(12\%), the larger number of photoexcited carriers relax in a
similar manner, except that the broad carrier distribution is centered at,
not below, the barrier energy (Fig.~\ref{MQW-DT}b). This suggests that the
QW states are saturated at high excitation densities and that excess
photoexcited carriers accumulate in the 3D barrier states. For both
excitations, the blue edge of the carrier distribution begins to decay
after 1 ps, but the red edge remains almost constant until 400 ps. The
rate of blue edge decay is significantly faster for the above I$_{th}$
excitation than below (Fig.~\ref{MQW-DT}b), explaining the pump
intensity-dependent behavior of the fast component in the SI-TRDT
(Fig.~\ref{IntegratedDT}a). It is apparent that during the SE process,
carriers cascade to refill the emission-emptied QW states from higher
energy 3D barrier states and 2D QW states. \cite{OzgurPRB,SatakePRB1999}
After SE ends, a much slower, pump intensity-independent decay and
redshift is observed as the carriers are lost through recombination. By
400 ps, the carrier distribution is centered at the QW band edge (3.11
eV). The decay constant for the TRDT signal at the QW energy (3.11 eV) is
0.66$\pm$0.06 ps, which agrees with the slow SI-TRDT rate and the
previously measured recombination time. \cite{OzgurAPL2000}

SR-TRDT data for the 23\% sample for excitation densities below
(Fig.~\ref{MQW-DT}c) and above (Fig.~\ref{MQW-DT}d) I$_{th}$(23\%)
presents a slightly different behavior. Since the localized and confined
QW states are deeper, initial carrier accumulation is not observed at the
barrier energy, and below I$_{th}$ excitation produces a broader (420 meV)
carrier distribution whose peak occurs near the QW band edge ($\sim$2.93
eV) in only 0.5 ps. The carriers arrive in the QWs (2.93 eV) 0.85 ps later
than at the barriers (3.29 eV), indicating the carrier capture process is
significantly slower than in the 12\% sample. By contrast, for the above
I$_{th}$(23\%) excitation, a wider and flatter carrier distribution is
observed in 1 ps, extending from the barriers to the localized states
below the QWs. As in the 12\% sample, the blue edge of the carrier
distribution is responsible for the fast component in
Fig.~\ref{IntegratedDT}b, which peaks in $<$1 ps and quickly decays while
SE is underway ($<$10 ps). Likewise, the carrier distribution peak remains
near the QW band edge, and carriers cascade from higher energy states to
refill the SE-emptied QW states. After SE ends, the distribution continues
to redshift slowly as carriers are lost through various decay channels.
The carriers at the QW band edge are observed to decay much faster
(0.32$\pm$0.05 ns at 2.90 eV) than in the 12\% sample. This occurs
because, in addition to radiative decay, QW carriers in the 23\% sample
decay non-radiatively into the localized states as much as 200 meV below
the QW band edge. Carrier decay is slower in these localized states
(0.48$\pm$0.08 ns at 2.84 eV) than at the QW band edge, suggesting that
the presence of these localized states undermines the QW radiative
emission process.  As further evidence of the role of inhomogeneities in
the 23\% sample, note the long-lived induced absorption at the barrier
energies, also due to carriers localized in traps after 10 ps.

For the 23\% sample, then, localization in material inhomogeneities plays
a greater role in the carrier redistribution and relaxation processes.  A
larger percentage of carriers are trapped by the inhomogeneities at the
barriers and in the localized states below the SE ($\sim$QW) energy,
making them energetically unavailable to participate in the refilling
process required to sustain SE. The increased importance of non-radiative
relaxation produces carrier decay times at the QW band edge that are
remarkably faster for the 23\% than for the 12\% sample. The combination
of these effects leads to a greater variety of carrier relaxation channels
and a commensurate increase of I$_{th}$(23\%) over I$_{th}$(12\%). These
observations reconcile the apparently conflicting findings of Kwon
\textit{et al.} and Yablonskii \textit{et al.} regarding the dependence of
I$_{th}$ on the QW depth.\cite{KwonAPL1999,YablonskiiMSEB2001} In the
former case, which concerns GaN/In$_x$Ga$_{1-x}$N MQWs with $x<13.3\%$,
$I_{th}$  is low, as reported here, because of the (reduced) material
inhomogeneities. However, $I_{th}$ was found to increase with decreasing
$x$ because of the increasing role of other non-radiative
pathways.\cite{KwonAPL1999}  In the latter case, emission from MQWs with
unknown but larger $x$ exhibits the same sort of increase in $I_{th}$ with
$x$ as reported here. They similarly conclude that non-radiative processes
involving (increased) inhomogeneities is
responsible.\cite{YablonskiiMSEB2001} The transition for $I_{th}$ behavior
appears to occur at a critical value for $x$ ($\sim$0.20), above which
inhomogeneities, Stokes-like shifts, and radiative decay times are known
to increase rapidly.\cite{ChichibuJVST1998}

This work was partially supported by ARO grant DAAG55-98-D-0002-0007.

\begin{references}
\bibitem{SEbelowTh}
Weak SE is still seen below I$_{th}$ due to inhomogeneities and the
nonuniform intensity profile of the pump spot.
\end{references}
\end{document}